%% file: main.tex
\documentclass[conference]{IEEEtran}
\IEEEoverridecommandlockouts
\usepackage{cite}
\usepackage{amsmath,amssymb,amsfonts}
\usepackage{algorithmic}
\usepackage{graphicx}
\usepackage{textcomp}
\usepackage{xcolor}
\def\BibTeX{{\rm B\kern-.05em{\sc i\kern-.025em b}\kern-.08em
    T\kern-.1667em\lower.7ex\hbox{E}\kern-.125emX}}
    
\usepackage{url}
\usepackage{amsmath}
\usepackage{siunitx}
\usepackage{pifont}
\newcommand{\cmark}{\ding{51}}%
\newcommand{\xmark}{\ding{55}}%
\usepackage{multirow}
\usepackage[nolist]{acronym}

\usepackage{booktabs}

\usepackage{float} 
\usepackage{caption}
\usepackage{subcaption}
\usepackage{verbatim}
\usepackage{hyperref}

\usepackage[resetlabels,labeled]{multibib}
\newcites{App}{Appendix References}

\mathchardef\mhyphen="2D
    
\begin{document}

\title{DESED-FL and URBAN-FL: Federated Learning Datasets for Sound Event Detection
}

% \author{\IEEEauthorblockN{David Johnson}
% \IEEEauthorblockA{\textit{Industrial Media Applications} \\
% \textit{Fraunhofer Institute for Digital Media Technology (IDMT)}\\
% Ilmenau, Germany \\
% \and
% \IEEEauthorblockN{Wolfgang Lorenz}
% \textit{Fraunhofer Institute for Digital Media Technology (IDMT)}\\
% Ilmenau, Germany \\
% \and
% \IEEEauthorblockN{Michael Taenzer}
% \textit{Fraunhofer Institute for Digital Media Technology (IDMT)}\\
% Ilmenau, Germany \\
% \and
% \IEEEauthorblockN{Stylianos Mimilakis}
% \textit{Fraunhofer Institute for Digital Media Technology (IDMT)}\\
% Ilmenau, Germany \\
% \and
% \IEEEauthorblockN{Jakob Abe{\ss}er}
% \textit{Fraunhofer Institute for Digital Media Technology (IDMT)}\\
% Ilmenau, Germany \\
% \and
% \IEEEauthorblockN{Sascha Grollmisch}
% \textit{Fraunhofer Institute for Digital Media Technology (IDMT)}\\
% Ilmenau, Germany \\
% \and
% \IEEEauthorblockN{Hanna Lukashevich}
% \textit{Fraunhofer Institute for Digital Media Technology (IDMT)}\\
% Ilmenau, Germany \\

\author{\IEEEauthorblockN{David S. Johnson\IEEEauthorrefmark{1}, Wolfgang Lorenz, 
Michael Taenzer, Stylianos Mimilakis, \\
Sascha Grollmisch, Jakob Abe{\ss}er and Hanna Lukashevich}
\IEEEauthorblockA{Fraunhofer Institute for Digital Media Technology (IDMT), Ilmenau, Germany}
\IEEEauthorblockA{\IEEEauthorrefmark{1}david.scott.johnson@idmt.fraunhofer.de}
}

\maketitle

\begin{abstract}
Research on \ac{AED} in environmental settings has seen increased attention in recent years. The large amounts of (private) domestic or urban audio data needed raise significant logistical and privacy concerns. The inherently distributed nature of these tasks, make \ac{FL} a promising approach to take advantage of large-scale data while mitigating privacy issues. While \ac{FL} has also seen increased attention recently, to the best of our knowledge there is no research towards \ac{FL} for \ac{AED}. To address this gap and foster further research in this field, we create and publish novel \ac{FL} datasets for \ac{AED} in domestic and urban environments. Furthermore, we conduct baseline results on the datasets in a \ac{FL} context for three deep neural network architectures. The results indicate that \ac{FL} is a promising approach for \ac{AED}, but faces challenges with divergent data distributions inherent to distributed client edge devices.  
\end{abstract}


\input{intro.tex}
\input{related_work.tex}
\input{datasets.tex}
\input{experiments.tex}
\input{results.tex}
\input{conclusion.tex}

\input{acronyms}

\bibliographystyle{IEEEtran}
\bibliography{IEEEabrv,literature.bib}

% COMMENT OUT THE LINES BELOW BEFORE SUBMITTING
% ALSO REMEMBER TO UPDATE APPENDIX FOOTNOTES
\onecolumn
\input{appendix.tex}

\bibliographystyleApp{IEEEtran}
\bibliographyApp{IEEEabrv,literature.bib}

\end{document}

%% file: intro.tex
\acresetall

\section{Introduction} \label{sec:intro}

The aim of \ac{AED} is to automatically identify the occurrence of target sound events, such as glass breaking or dog barking, within an audio signal capturing an acoustic scene. 
% Typically, sound events in an acoustic scene have a wide variety of characteristics, ranging from short and transient (i.e. a gun shot) to long and stationary (i.e. an air conditioner), and may overlap with each other, yielding a polyphonic mixture of audio events combined with varying types of background noises.
Identifying these sound events within complex scenes is a challenging and open research problem that has attracted much attention in recent years, observed with the increase in the literature and particularly with the growing research interest in the DCASE Community\footnote{\url{http://dcase.community}}.
Two common \ac{AED} uses-cases are acoustic monitoring in domestic \cite{serizel2020:desed} and urban \cite{salamon14:urbansounds} environments.
In both scenarios, the use of audio to train detection models in a centralized training context raises considerable privacy concerns. In these environments, there are speech and other sounds that are confidential which should not be shared or stored insecurely.
% Urban environments face similar privacy concerns exacerbated by the fact that most individuals have not given direct consent to be recorded. 
% These concerns are increasingly gaining attention as real-world privacy issues come to light with reports of companies sharing customer recordings with employees for human review\footnote{\url{https://www.theguardian.com/technology/2019/apr/11/amazon-staff-listen-to-customers-alexa-recordings-report-says} (last accessed: 02 Feb. 2021)}. 
% Consequently, \ac{AED} systems must implement privacy-by-design approaches to gain users' trust~\cite{lau2018alexa}.

State-of-the-art approaches to \ac{AED} are most commonly based on deep learning \cite{dang2017:aedsurvey} which requires large centralized datasets for model training, posing significant security and logistical challenges. %that grow increasingly large with added users.
% Considering the previous data privacy concerns, %organizations need to provide guarantees regarding the security of their users' data.
\Ac{FL}~\cite{McMahan_Fl_2017} offers an attractive approach to mitigate some of these concerns. % and storage
Instead of sending private data to a centralized data store, \ac{FL} performs model training directly on many client edge devices (from here on referred to as clients) using locally stored data. The clients then share only their updated parameters with a coordination server, which aggregates the shared parameters to update a global model. The new global model is then transferred back to the clients. This process continues until convergence, or indefinitely if new data is continuously acquired~\cite{Kairouz_FlReport_2019}. 
%Consequently, each edge device is transferred only the aggregated weights from multiple private data collections, without the necessity of sharing the actual data. It should be noted that the interaction between the centralized coordination server and the edge devices is taking place over secure protocols to enhance the overall security and privacy~\cite{McMahan_Fl_2017}.

%In typical centralized context, datasets for training and evaluation are typically assumed to be \ac{IID}. In researching \ac{FL} contexts, however non-\ac{IID} datasets~\cite{McMahan_Fl_2017, Sattler2019.noniid, hsieh2019non} are needed, to account for the distributed nature of \ac{FL}. These can introduce different background and acoustic characteristics, as well as the availability of sound events at different locations.
%Local models tend to fit to the distribution contained in the local data, which has been shown to negatively impact the performance of aggregated models trained on non-\ac{IID} data~\cite{hsieh2019non, Sattler2019.noniid, McMahan_Fl_2017}.
%Non-\ac{IID} data is already a common problem for \ac{AED} due to variations in background environments, recording devices, and room acoustics \cite{abesser2020:aedsurvey}. 

Current research in \ac{FL} has focused on image or text-based tasks. To our knowledge, the only known research or practical applications of \ac{FL} in the audio domain are related to keyword spotting \cite{Leroy2019.keyword, apple2019.heysiri, bhowmick2018.protection}. Due to the limited research on \ac{FL} for \ac{AED}, there remain questions about the effectiveness of the approach due to varying acoustic conditions inherent to distributed clients. For example, data may be captured from clients in multiple locations with different background noise characteristics or in locations with only a subset of the sound event classes. This leads to differences in data distributions amongst the clients involved in the training process. For centralized training, data from multiple devices is combined into single training dataset that is typically assumed to be \ac{IID}, but with \ac{FL} distributed data collection leads to models being trained using data from divergent distributions, i.e., data that is non-\ac{IID}. Existing \ac{AED} datasets do not capture the non-\ac{IID} characteristics seen with \ac{FL}. To address this gap and foster research on \ac{FL} for \ac{AED}, we contribute novel \ac{AED} datasets specifically designed for \ac{FL} training.
% using source material from the existing \ac{AED} datasets, \ac{URBAN-8K} \cite{salamon14:urbansounds} and \ac{DESED} \cite{serizel2020:desed}. 
Additionally, we provide baseline results for three neural network architectures to evaluate the effects of \ac{FL} hyperparameters and non-\ac{IID} data on \ac{AED} performance.

%% file: related_work.tex
\begin{table*}[h]
\renewcommand{\arraystretch}{1.25}
\caption{Sound event and background classes for each of the datasets.}
\label{tab:classes}
\centering
\begin{tabular}{@{} l p{0.55\linewidth} p{0.3\linewidth} @{}} \toprule
\textbf{Dataset} & \textbf{Sound Event Classes} & \textbf{Background Noise Classes} \\\midrule

 DESED-FL & $e_1$:~Dishes; $e_2$:~Cat; $e_3$:~Frying; $e_4$:~Dog; $e_5$:~Blender; $e_6$:~Speech; $e_7$:~Vacuum cleaner; $e_8$:~Electric shaver/toothbrush; $e_9$:~Alarm bell; $e_{10}$:~Running water & apartment room, computer interior, computer lab, emergency staircase, and library \\
 
 URBAN-FL & $e_1$:~Children playing; $e_2$:~Siren; $e_3$:~Drilling; $e_4$:~Street music; $e_5$:~Car horn; $e_6$:~Gun shot; $e_7$:~Jackhammer; $e_8$:~Dog bark; $e_9$:~Air conditioner; $e_{10}$:~Engine idling & birds, crowd, fountain, rain, and traffic \\

\bottomrule
\end{tabular}
\end{table*}

\section{Related Work} \label{sec:rw}

% Both \ac{FL} and \ac{AED} have seen an increase in research in recent years.  In this section, we provide a brief discussion of the relevant literature for each topic.  

\subsection{Federated Learning}
McMahan et al. first proposed the idea of \ac{FL} as a method to allow data to remain on distributed devices while training a shared model by aggregating locally trained updates~\cite{McMahan_Fl_2017}. For a comprehensive overview of \ac{FL}, refer to the technical report by Kairouz et al.~\cite{Kairouz_FlReport_2019}. 
% Since the state-of-the-art approaches for AED rely on deep learning, this subsection focuses on \ac{FL} related to deep neural networks.
%
There are two main challenges for \ac{FL} methods. First is the need for communicating over unreliable networks to transmit data. A second issue results from data captured in varying contexts, leading to datasets that are statistically dissimilar, non-\ac{IID}, between clients.  

To address these challenges, %and improve performance in real-world applications, 
Sattler et al.~\cite{Sattler_FlCommEfficient_2019} proposed a compression framework \ac{STC}. Similarly, Lin et al.~\cite{Lin_DGC_2017} and Bernstein et al.\cite{Bernstein_signSGD_2018} proposed methods to remove the redundancies of gradient information in node-distributed learning frameworks.
% like \ac{FL} by sparsifying the gradients and preserving only the relevant information~\cite{Lin_DGC_2017} or transmitting only the sign~(-1, 0, or~1) of the gradients~\cite{Bernstein_signSGD_2018}. 
Hsieh et al. evaluated the challenges of non-\ac{IID} data in an \ac{FL} scenario \cite{hsieh2019non}. They identified problems with the batch normalization layer, a common layer in many \ac{DNN} architectures, and proposed to use group normalization~\cite{Wu2018.groupnorm} instead. Similarly, to address the problems of  non-\ac{IID} data, Sattler et al.~\cite{Sattler_ClusteredFl_2019} proposed % to exploit the geometric surface of the loss function that the centralized server has to minimize. This is done by performing 
a clustering operation to group clients whose data distributions have similar characteristics. While the previous research proposed methods to overcome challenges in \ac{FL}, there are no known datasets to evaluate them for \ac{AED}. We address this gap, by presenting new \ac{AED} datasets specifically designed for \ac{FL} with non-\ac{IID} data.

% Another improvement to the typical federated averaging is proposed in~\cite{Goetz_ActiveFL_2019}. Specifically, it is proposed to replace the common average across edge devices, or the random selection of edge devices by a selective process. That process takes into account the confidence of each model in each edge device. Then, the models with the highest confidence are used to compute the average, essentially maximizing the efficiency in training and performance.

\subsection{Sound Event Detection}
State-of-the-art \ac{AED} algorithms build upon deep neural networks, the most common being \ac{CNN} and \ac{CRNN} based architectures.
Both architectures include convolutional front-ends, where multiple convolutional layers are trained to learn sound-specific features.
As input to the network, either fixed two-dimensional signal transformations such as mel spectrograms \cite{Hou:2019:SoundEvent:ICASSP} or raw one-dimensional audio samples are used (end-to-end learning) \cite{Cakir:2018:EndtoEndAED:IJCNN}.
As a back-end, \acp{CNN} use fully-connected layers for classification whereas \acp{CRNN} employ recurrent layers such as \ac{GRU} or \ac{LSTM} layers to model the temporal progression of the extracted features. We focus our work on \ac{CNN} architectures for a lightweight approach, to enable model training on low resource devices required by \ac{FL}.

Training \ac{AED} models requires strongly labeled datasets in which onset and offset times are labeled for each sound event. Because of the laborious effort required to annotate real-world samples, researchers often use synthetically generated datasets. This requires mixing events from a curated sound bank with a background signal to synthesize soundscapes with multiple, possibly overlapping, events. For example, URBAN-SED~\cite{Salamon2017} is composed of sound events from the \ac{URBAN-8K}~\cite{salamon14:urbansounds} dataset mixed with Brownian noise. A recent trend is to use a combination of synthetic and real recordings for training and evaluation as with \ac{DESED}~\cite{serizel2020:desed}. For both datasets, however, events are distributed uniformly during soundscape generation. For \ac{FL}, events should be distributed in a structured fashion to simulate real-world distributed learning conditions. To enable research in \ac{FL} for \ac{AED}, our proposed datasets distribute soundscapes to clients with different background characteristics and class distributions.

%% file: datasets.tex
\section{Datasets}\label{sec:datasets}

In this section we present DESED-FL and URBAN-FL, datasets for acoustic monitoring of domestic and urban environments with \ac{FL}. Each use case contains two independent training sets: an \ac{IID} dataset, in which sound event classes are distributed evenly to devices, and a non-\ac{IID} dataset, in which only a subset of the total classes is assigned to each client.
% It is expected that \ac{AED} edge devices will be not as widespread as other devices used for \ac{FL}, such as smart phones; therefore, 
To imitate varying acoustic conditions, we mix sound events with one of five background noise classes. Each training dataset is partitioned into 100 clients with 20 clients per background noise class. It is possible to simulate more than 100 clients by partitioning further, or less than 100 by combining or removing clients. For reproducible evaluation, each use case also includes an evaluation dataset in which sound events are uniformly distributed to each background class.

To generate the soundscapes for DESED-FL, sound events and background noises are sourced from \ac{DESED}~\cite{serizel2020:desed}. URBAN-FL soundscapes are generated using sound events from \ac{URBAN-8K}~\cite{salamon14:urbansounds}, and background noises from \ac{IUSD}~\cite{gloaguen2018.iusd}. The sound event and noise classes for each dataset are listed in Table \ref{tab:classes}.

% Sound event classes from \ac{DESED} include alarm/bell ringing, blender, cat, dishes, dog, electric shaver/toothbrush, frying, running water, speech, and vacuum cleaner.
% Additionally, the dataset contains five types of background noise, apartment room, computer interior, computer lab, emergency staircase, and library.
% , which includes the classes air conditioner, car horn, children playing, dog bark, drilling, engine idling, gun shot, jackhammer, siren, and street music. For background noise we select five types from the \ac{IUSD}\footnote{\url{https://zenodo.org/record/1213793}} that minimize the number of sounds similar to those in 10 sound event classes, including birds, crowd, fountain, rain, and traffic.
% Each dataset comprises 10 event classes and 5 background noise classes. The task is to detect the events which are placed over the backgrounds. The procedure for this is described in the following.

%The usage for both datasets should be the same.
% The federated learning experiments use up to 100 different edge devices, which learn to detect foreground events. To emulate different environments, the events are evenly distributed over the 5 background noises. In the case of 100 edge devices, this results in 20 edge devices per background noise.

\subsection{\ac{IID} and Non-\ac{IID} Training Datasets}\label{subsec:datasets:training_sets}

% \begin{table}[ht]
%     \centering
%     \resizebox{\columnwidth}{!}{
%     \begin{tabular}{r|ccccc}
%         1 & Dishes & Cat & Dog & Speech & Alarm bell ringing \\
%         2 & Dishes & Frying & Dog & Vacuum cleaner & Electric shaver/toothbrush \\
%         3 & Dishes & Frying & Blender & Speech & Running water \\
%         4 & Cat & Frying & Blender & Vacuum cleaner & Alarm bell ringing \\
%         5 & Cat & Dog & Blender & Electric shaver/toothbrush & Running water
%     \end{tabular}}
%     \caption{Non-\ac{IID} distribution used for the \ac{DESED}-FL dataset.}
%     \label{tab:desed_bad_dist}
% \end{table}

% \begin{table}[ht]
%     \centering
%     \resizebox{\columnwidth}{!}{
%     \begin{tabular}{r|ccccc}
%         1 & Children playing & Siren & Street music & Gun shot & Air conditioner \\
%         2 & Children playing & Drilling & Street music & Jackhammer & Dog bark \\
%         3 & Children playing & Drilling & Car horn & Gun shot & Engine idling \\
%         4 & Siren & Drilling & Car horn & Jackhammer & Air conditioner \\
%         5 & Siren & Street music & Car horn & Dog bark & Engine idling
%     \end{tabular}}
%     \caption{Non-\ac{IID} distribution used for the \ac{URBAN}-FL dataset.}
%     \label{tab:urban_bad_dist}
% \end{table}

% To acquire a better understanding of how real-world data distributions affect the performance of models trained through \ac{FL}
To generate the \ac{IID} and non-\ac{IID} dataset variations, two sound event class distribution schemes are implemented for assigning events to clients. 
The \ac{IID} scheme uniformly distributes event classes to all devices. This is the best possible case for training \ac{FL} models since all clients have access to all classes.
To simulate a more realistic scenario, in which clients only have knowledge of a subset of all classes, the non-\ac{IID} scheme distributes events to clients using five class distributions, with each distribution containing a subset of five classes. 
% For the non-\ac{IID} scheme, sound event classes are distributed to the clients using one of five class distributions each containing a subset of five classes.
To minimize the bias that could result from linking background sounds to certain subsets of event classes, each of the five distributions is assigned to four clients per background noise. This results in each set of 20 clients per background noise comprising five class distributions. A detailed view of each data distribution is presented in Appendix 
\ref{app:data}.
% A\footnote{\label{fn:app}\url{http://arxiv.org/abs/2102.08833}}.
% See Figures \ref{subfig:desednoniid} and \ref{subfig:urbannoniid} for the distribution of data among each device by event duration. Here, for each background, you can see five clear groups of devices with common distributions of events.
In the non-\ac{IID} scheme, each of the five class distributions contains five event classes for a total of 25 event classes, which is not divisible by 10, the number of total classes. This means that five event classes were used three times and five event classes were used only twice. 

One goal in designing the distributions is to have each set of classes be as different as possible from any other set, while ensuring that each class is used at least once. Hence, the algorithm to select class distributions minimizes the penalty value $p$ computed by \[
    p = \sum_{k=1}^{N_\text{coll}-1} \sum_{l=k+1}^{N_\text{coll}} N_\text{eq}^2\left(k, l\right),
\] where $N_\text{coll}$ equals the number of collections and $N_\text{eq}\left(k, l\right)$ represents the number of equal classes in collections $k$ and $l$. We do not take into account whether classes are likely to occur together in real-world applications, since there are no obvious combinations that would never happen together.   

The distributions that minimize the penalty $p$ consists of the sets $\left\{ e_1, e_2, e_4, e_6, e_9 \right\}$, $\left\{ e_1, e_3, e_4, e_7, e_8 \right\}$, $\left\{ e_1, e_3, e_5, e_6, e_{10} \right\}$, $\left\{ e_2, e_3, e_5, e_7, e_9 \right\}$, and~$\left\{ e_2, e_4, e_5, e_8, e_{10} \right\}$, where each event $e_i$ is assigned one of the available event classes. The mapping of the event classes to the positions (i.e. $e_\#)$ have been randomized, and can be found in Table \ref{tab:classes}. 
% For \ac{DESED} this is $e_1$:~Dishes; $e_2$:~Cat; $e_3$:~Frying; $e_4$:~Dog; $e_5$:~Blender; $e_6$:~Speech; $e_7$:~Vacuum Cleaner; $e_8$:~Electric shaver/toothbrush; $e_9$:~Alarm bell; $e_{10}$:~Running water. For \ac{URBAN-8K} this is $e_1$:~Children playing; $e_2$:~Siren; $e_3$:~Drilling; $e_4$:~Street music; $e_5$:~Car horn; $e_6$:~Gun shot; $e_7$:~Jackhammer; $e_8$:~Dog bark; $e_9$:~Air conditioner; $e_{10}$:~Engine idling.

%Tables~\ref{tab:desed_bad_dist} and~\ref{tab:urban_bad_dist} show all collections with respect to the restrictions for the non-\ac{IID} distribution.

\subsection{Data Generation}\label{subsec:datasets:scaper}

The datasets consist of ten-second soundscapes synthetically generated using Scaper~\cite{Salamon2017} by mixing between one and five possibly overlapping source events with one background noise type. Each event is mixed with an \ac{SNR} chosen from $\mathcal{N}(\mu,\,\sigma^{2})$ with $\mu=\si{10}{dB}$ and $\sigma=\si{3}{dB}$.
The sound events are selected by sampling from the class distributions as discussed in Section \ref{subsec:datasets:training_sets}. Additionally, source events are augmented by pitch shifting the audio by an amount uniformly sampled from the range $[-2, 2]$, and by time stretching by a value uniformly sampled from the range $[0.8, 1.25]$. These augmentations are only applied to the training data.

Before generating the soundscapes, the source data is split into training and evaluation sets to ensure that there is no data leakage. For DESED-FL events, the data is partitioned into training and evaluation data according to split provided by \ac{DESED}.
For Urban-FL, we take the approach employed for URBAN-SED, by using the existing stratified folds from the \ac{URBAN-8K} dataset for the split: folds 1-6 are used for training and 9-10 for evaluation. 
We omit folds 7-8, used by URBAN-SED for validation, since the DESED source data does not contain a validation split. 
The background noise for each dataset is split into training and evaluation by splitting each source file into separate training and evaluation segments.

The final training datasets each contain 100 ten-second soundscapes per edge device, totaling \num{10000} soundscapes. The evaluation datasets contain \num{400} soundscapes per background class for a total of \num{2000} soundscapes. To enable reproducibility, the dataset creation scripts are available for download\footnote{\url{https://www.idmt.fraunhofer.de/en/publications/idmt-fl.html}}

% Each of the background noise classes in \ac{DESED} consists of one long audio file (between 44 and 68 minutes). This makes a file-based separation impossible here.

% The event classes in the raw material of the \ac{DESED} data already come divided into a training and evaluation set, which we use for \ac{DESED}-FL.
% For URBAN-FL, this split is done manually to generate an evaluation set that only contains material unknown to the trained model. We use a similar split ratio per class as the original \ac{DESED} split (approx. \SI{77}{\percent} / \SI{23}{\percent}).

%% file: experiments.tex
\section{Experimental Setup} \label{sec:exps}

\subsection{Architectures}\label{sec:exps:archs}

We propose three baseline architectures to evaluate different model complexities and their effects on \ac{FL}. One of the goals driving this research is developing small models that are able to be trained on low resource devices, such as neuromorphic hardware. Therefore, we evaluate two standard \ac{CNN} architectures of different sizes, and a \ac{ResNet} architecture~\cite{Koutini2019}. The baseline \ac{CNN}, \emph{CNN-Base}, is a medium-sized architecture based on the feature extraction front-end of the 2019 \ac{DCASE} Challenge Task 3 baseline architecture. It is composed of seven convolutional blocks and a linear classification layer, for a total of \num{542442} parameters. The second \ac{CNN} architecture is a small \ac{CNN}, called \emph{CNN-Sm}, and was designed using neural architecture search with Bayes optimization~\cite{Snoek2012:bayes_opt} to limit the model to nearly \num{100000} parameters while optimizing the F-score on the URBAN-SED dataset. The found model has four convolutional blocks, a single feed forward layer, and a classification layer, resulting in \num{115434} parameters. Lastly, we propose a medium-sized \ac{ResNet} architecture, \emph{ResNet}, with five \ac{IC} \ac{ResNet} blocks~\cite{chen2019.rethinking} and a classification layer for a total of \num{422090} parameters. Detailed descriptions for each architecture may be found in Appendix 
\ref{app:archs}. %uncomment for arxiv
% B\footref{fn:app}. %uncomment for eusipco submission

\subsection{Preprocessing}
\subsubsection{Input Representation}
% The input representation was inspired by the work of Koutini et~al.~\cite{Koutini2019} in their work to regularize \acp{CNN} for acoustic scene classification by minimizing the receptive field. 
The input for each model is a perceptually weighted mel spectrogram~\cite{Koutini2019}: The input signal is first downsampled to \SI{22050}{\hertz}. The \ac{STFT} is applied with a \ac{FFT} of size \num{2048} and a hop size of \num{512}, and is followed by perceptual weighting. A mel-filter bank of 256 mel bands is then applied. Finally, 43~windows are stacked together resulting in a feature representation (43x256x1) of one second. 

\subsubsection{Data Augmentation}
Similar to Salamon and Bello~\cite{Salamon2017:dataaug}, we apply pitch shifting to the audio data before extracting the mel spectrograms.  However, instead of applying all shifts of $\pm2$ and $\pm1$ semitones for a total increase factor of 4, we randomly select one semitone value between $\pm2$ for each input for a total augmentation factor of 1 (i.\~e., doubling the size of the dataset). This helps significantly reduce the final size of the dataset and has only a minor impact on model performance.

\subsection{Experimental Design}

In our first experiment, the three proposed network architectures are evaluated in a centralized training scenario. First, we train the models using the original URBAN-SED dataset~\cite{salamon14:urbansounds} to validate that the architectures work as expected on a well-known dataset. Then, we train each of the three models using the \ac{FL} training sets. We evaluate the models, first, with batch normalization, and then, replace all batch normalization layers with group normalization to ensure that this substitution for mitigating non-\ac{IID} issues in \ac{FL}~\cite{hsieh2019non} does not significantly affect the baseline results. 
All models are trained for 50 epochs with early stopping using a patience of 25 epochs, which monitors the validation loss. We use the Adam optimizer~\cite{kingma2014.adam} with a cosine learning rate schedule in the range $[1e-04, 1e-06]$.

The next experiments evaluate the influence of \ac{FL} hyperparameters on \ac{AED} performance, namely the total number of clients, $N$, the percentage of total clients that participate in each training round (participation rate $r_p$), and number of local epochs performed during each round, $E_L$. For $N$ we evaluate 1, 25, and 100 clients, with $N=1$ providing a \ac{FL} baseline. For $r_p$, values of $\{0.25, 0.5, 1.0\}$ are used in order to simulate the uncertainty of client participation during each communication round. At each round a fraction, $r_p$, of the $N$ clients are uniformly sampled (with replacement at the end of each round) to participate in training.
As the amount of network communication should be limited in \ac{FL}, we evaluate the effects of $E_L=\{1, 3, 5\}$ to reduce the number of communication rounds by increasing amount of training steps performed on each client before sending weight updates. For all \ac{FL} experiments, the local client models are optimized using Adam, with a learning rate $lr=1e-04$, proposed by Leroy et al. \cite{Leroy2019.keyword}. The local weight updates are aggregated by the coordinator using the standard Federating Averaging algorithm using \ac{SGD} with a learning rate of $1.0$~\cite{McMahan_Fl_2017}. 

Due to the number of experiments, the size of the datasets, and the training times required for each experiment, we limit the number of communication rounds during training to 60 rounds per experiment. Thus, reducing the time and resources required for evaluation\footnote{In real-world \ac{FL}, this would not be an issue due to the inherent parallelization of \ac{FL}, as opposed to simulating \ac{FL} on a single server.}.  While in some cases the models may not have completely converged, the results provide valuable insights to improve our understanding of the different effects of \ac{FL} hyperparameters to better focus future research.

%% file: results.tex
\section{Results}\label{sec:results}

% In this section, we discuss the results of the experiments on the proposed datasets.  First, in Section \ref{sub:results:central}, we discuss the the results of the centralized learning scenario to provide a baseline.  Then the \ac{FL} results are presented in Section \ref{sub:results:fl}. 

\subsection{Centralized Training}\label{sub:results:central}

Table \ref{tab:baseline} lists the segment-based F-Scores for each of the proposed datasets and architectures from centralized training. Additionally, we include the results of the architectures trained using the URBAN-SED~\cite{Salamon2017} dataset as a baseline to validate the architectures' performance on a well-established dataset.

In the case of URBAN-SED, the models perform comparable or better than the original baseline of $F=0.57$ from Salamon et al.~\cite{Salamon2017}. In general, the \ac{ResNet} architecture is typically the best performing model. Furthermore, replacing batch normalization with group normalization has only minor effects on the performance of all architectures, and in many of the cases it improves performance. The results also indicate that data distribution, either \ac{IID} or non-\ac{IID}, has minimal effect on model performance in a centralized context. This is expected since all data is used during each training epoch, and the model does not fit to a particular data distribution.

\begin{table}[t]
\renewcommand{\arraystretch}{1.15}
\caption{Centralized Training F-scores.}
\label{tab:central}
\centering
\resizebox{1.0\linewidth}{!}{%
\begin{tabular}{@{} l l l c c c @{}} \toprule
\textbf{Dataset} & \textbf{IID} & \textbf{Norm} & \textbf{CNN-Sm} & \textbf{CNN-Base} & \textbf{ResNet} \\\midrule

URBAN-SED & - & batch & 0.566 & 0.567 & \textbf{0.601} \\ 
URBAN-SED & - & group & 0.532 & 0.587 & 0.589 \\ \midrule \midrule

URBAN-FL & \cmark & batch & 0.600 & 0.625 & \textbf{0.647}  \\ 
URBAN-FL & \cmark & group & 0.574 & 0.638 & \textbf{0.647} \\ \midrule
URBAN-FL & \xmark & batch & 0.593 & 0.609 & 0.634 \\ 
URBAN-FL & \xmark & group & 0.564 & 0.618 & \textbf{0.642} \\ \midrule \midrule

DESED-FL & \cmark & batch & 0.627 & 0.632 & 0.630 \\ 
DESED-FL & \cmark & group & 0.628 & 0.632 & \textbf{0.664} \\ \midrule
DESED-FL & \xmark & batch & 0.618 & 0.625 & 0.625 \\ 
DESED-FL & \xmark & group & 0.621 & 0.634 & \textbf{0.647} \\ \bottomrule
\end{tabular}}
\label{tab:baseline}
\end{table}

\subsection{Federated Learning}\label{sub:results:fl}

\begin{figure*}[t]
\centering
\begin{subfigure}[b]{0.495\textwidth}
    \includegraphics[width=\linewidth]{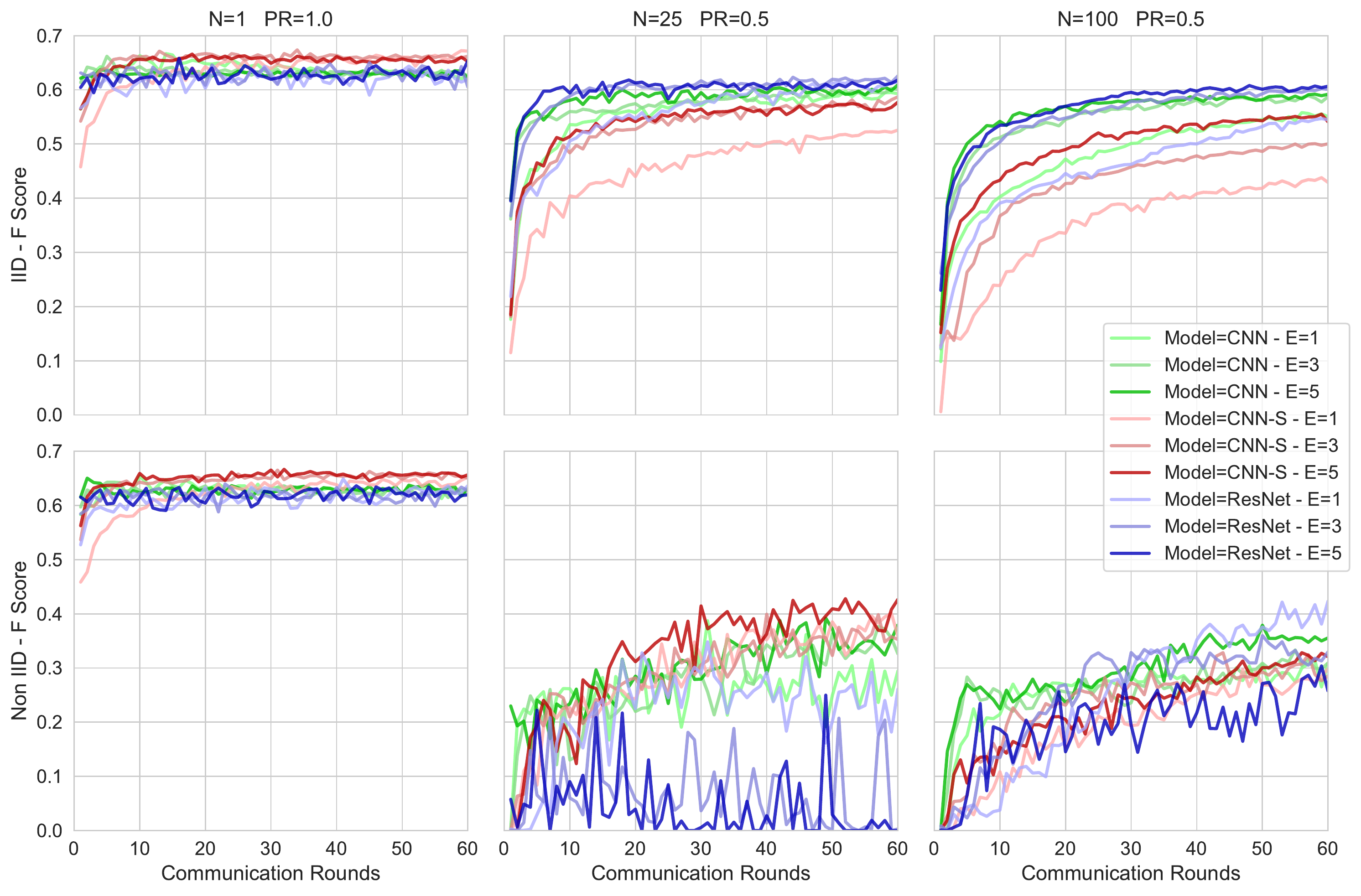}
    \caption{DESED-FL}
    \label{fig:desed:results_pr}
\end{subfigure}
\begin{subfigure}[b]{0.495\textwidth}
    \includegraphics[width=\linewidth]{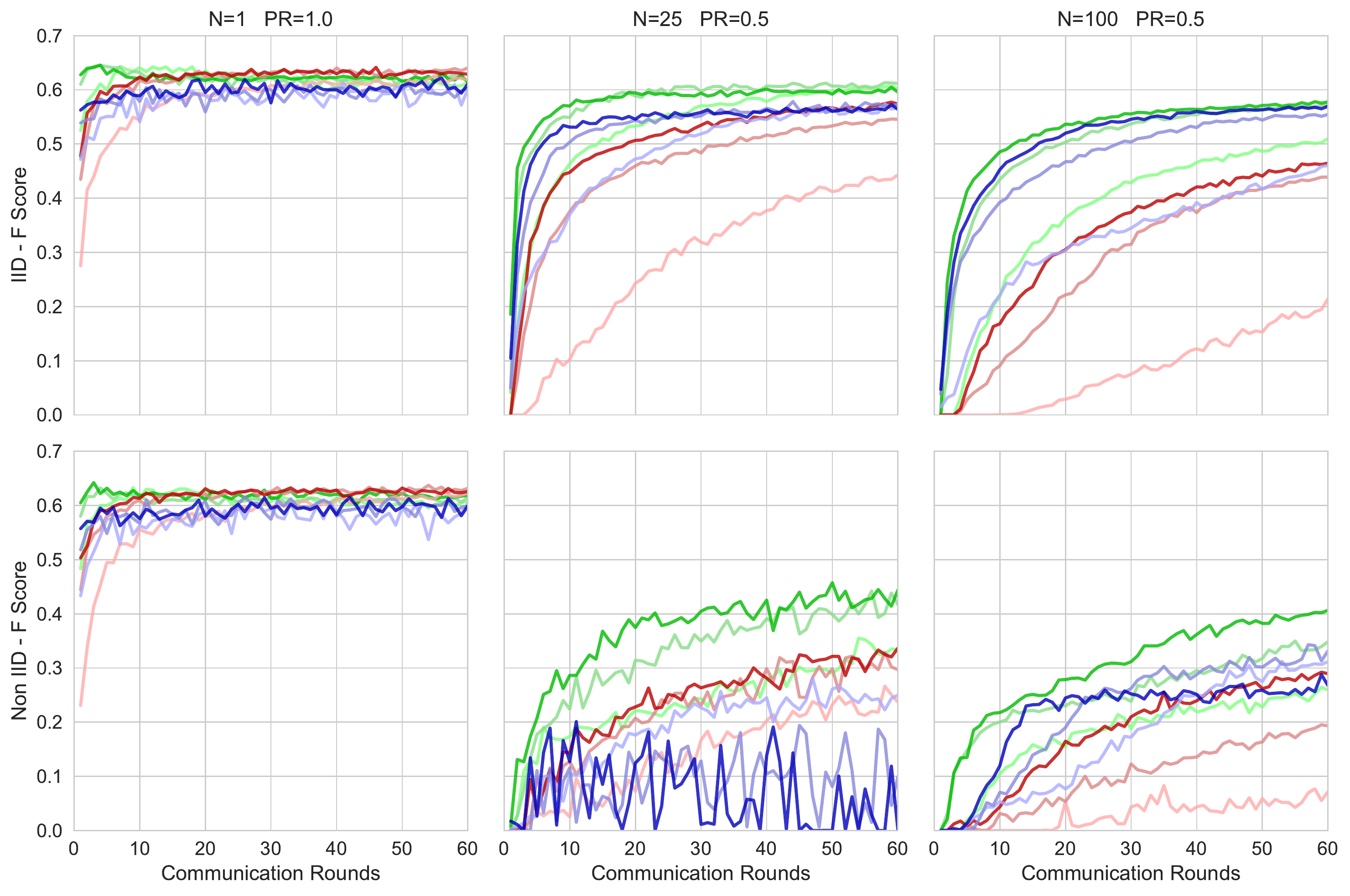}
    \caption{URBAN-FL}
    \label{fig:urban:results_pr}
\end{subfigure}
\caption{F-scores as a function of communication round for $PR=\SI{50}{\percent}$, and for $N=\{1, 25, 100\}$. The top row shows results of the \ac{IID} datasets and bottom row non-\ac{IID}. Each column in the subfigures show the results for a specified $N$.}
\label{fig:results}
\vspace{-0.2cm}
\end{figure*}

Figure \ref{fig:results} shows the segment-based F-scores on evaluation data for models trained with each training datasets. DESED-FL results are shown in Figure \ref{fig:desed:results_pr}, and URBAN-FL in Figure \ref{fig:urban:results_pr}. Each subfigure includes the the results from \ac{IID} (top row) and non-\ac{IID} (bottom row) datasets. The results are shown for hyperparameters $N$, $E_L$, and $r_p$ as a function of communication rounds.
Here we present the results for $r_p=0.5$ with the remaining results found in Appendix 
\ref{app:res}.
% C\footref{fn:app}.
The training curves for \ac{IID} data look similar to what is expected in a centralized training scenario with the training times (i.~e., number of communication rounds) being influenced by the architecture type, participation rate, and the number of local epochs. 
% We expect (based on empirical evidence) that with more training rounds each model would reach performance near the centralized baseline, but leave that for future research. 
Typically in the \ac{IID} setting, increasing model complexity and $E_L$ reduces the amount of communication needed between client and server by improving the training time; whereas, increasing $N$ slows down training. This could be attributed to the fact that each local client has less data resulting in smaller gradient deltas between the global and newly trained local model at each round. These observations, however, do not necessarily hold in a non-\ac{IID} context, where performance degrades significantly for all models (except for the baseline $N=1$). In this case, the results indicate that larger models are prone to overfit to the data seen during an individual training round.  This is especially true with a small number of clients, such as $N=25$, since there is inherently less variation in the randomly selected client distributions. However, even in cases when all data is seen during each communication round, i.~e., $r_p=1.0$, the \ac{ResNet} continues to overfit. Reducing $E_L$ helps to alleviate this issue by limiting the gradient values, but slowing down training. Training with a larger number of overall clients helps to mitigate this issue as well. In general, the models in a non-\ac{IID} scenario tend to overfit to local distributions, and the federating averaging process does not correct for this on its own. Damping the gradients before performing aggregation, by lowering the server learning rate or by normalizing gradients for example, may help to reduce the effects of large local gradients.

%% file: conclusion.tex
\section{Conclusion}\label{sec:conclusion}

In this work, we introduce DESED-FL and URBAN-FL, two novel datasets to foster research in \ac{FL} for \ac{AED}.
To better understand the effects of previously identified challenges associated with non-\ac{IID} data in \ac{FL}, we include both \ac{IID} and non-\ac{IID} training sets for each use case. 
Additionally, we contribute the first known research on \ac{FL} for \ac{AED} through the evaluation of three baseline neural network architecture. The results show that while \ac{FL} is a promising approach for \ac{AED} it 
is prone to challenges with non-\ac{IID} data similar to previous \ac{FL} research~\cite{hsieh2019non, Sattler_FlCommEfficient_2019}. By contributing non-\ac{IID} datasets, we hope enable further research to identify potential solutions to mitigate these issues. 

Future research directions include addressing the issues of non-\ac{IID} data through adaptive techniques such as the automatic adjustment of parameters~\cite{hsieh2019non}, or clustering techniques to identify groups of distributions and train individual models accordingly~\cite{Sattler_ClusteredFl_2019}. Furthermore, an investigation on strategies for local data management is needed. Since clients may have limited storage capacities, an evaluation of how much data should be stored and for how long is needed. Additionally, it is important to be able to identify which data samples contain the most information and should be stored for later training. To that end, the selective sampling of data based on entropy~\cite{wiewel2020.sampleselection} is an interesting research direction.

%% file: acronyms.tex
\begin{acronym}[URBAN-SED]
\acro{AED}[SED]{sound event detection}
\acro{CNN}{convolutional neural network}
\acro{CNN-S}{small \ac{CNN}}
\acro{CRNN}{convolutional recurrent neural network}
\acro{DCASE}{Detection and Classification of Acoustic Scenes and Events}
\acro{DESED}{Domestic Environment Sound Event Detection}
\acro{DNN}{deep neural network}
\acro{FFT}{fast Fourier transform}
\acro{FL}{federated learning}
\acro{GRU}{gated recurrent unit}
\acro{IC}{independent component}
\acro{IID}{independent and identically distributed}
\acro{LSTM}{long short-term memory}
\acro{ReLU}{rectified linear unit}
\acro{ResNet}{Residual Network}
\acro{RNN}{recurrent neural network}
\acro{SGD}{stochastic gradient descent}
\acro{STC}{sparse ternary compression}
\acro{STFT}{short-time Fourier transform}
\acro{SNR}{signal-to-noise ratio}
\acro{URBAN-8K}{UrbanSound-8K}
\acro{IUSD}[IUSD]{Isolated Urban Sound Database}
\end{acronym}

%% file: appendix.tex
\begin{appendices}

\section{Datasets}\label{app:data}

\begin{table}[ht]
\renewcommand{\arraystretch}{1.05}
\caption{DESED-FL and URBAN-FL Sound Event Statistics per Client Device}
\label{tab:stats}
\centering
\begin{tabular}{@{} l l c c c c c c @{}} \toprule
\textbf{Dataset} & \textbf{Version} & \textbf{Max Poly} & \textbf{Avg Poly} & \textbf{Avg Num Events} & \textbf{Avg Dur. (s)} & \textbf{Min Dur. (s)} & \textbf{Max Dur. (s)} \\ \midrule
URBAN-FL & \acs{IID} & 5 & $1.97\pm0.09$ & $2.98\pm0.14$ & $2.57\pm0.10$ & $0.09\pm0.03$ & $4.98\pm0.02$  \\
URBAN-FL & Non-\acs{IID} & 5 & $1.98\pm0.08$ & $3.00\pm0.14$ & $2.58\pm0.15$ & $0.09\pm0.03$ & $4.98\pm0.02$ \\ 
URBAN-FL & Eval & 5 & $1.98\pm0.05$ & $3.03\pm 0.08$ & $2.52\pm0.03$ & $0.11\pm0.00$ & $4.00\pm0.00$  \\ \midrule
DESED-FL & \acs{IID}  & 5 & $2.03\pm0.09$ & $2.98\pm0.14$ & $3.21\pm0.20$ &  $0.07\pm0.02$ & $10.0\pm0.00$ \\
DESED-FL & Non-\acs{IID}  & 5 & $1.99\pm0.28$ & $2.99\pm0.15$ & $3.08\pm1.18$ & $0.08\pm0.02$ & $9.65\pm1.16$ \\ 
DESED-FL & Eval & 5 & $2.02\pm0.04$ & $3.01\pm0.07$ & $2.99\pm0.04$ & $0.09\pm0.00$ & $10.0\pm0.00$  \\
\bottomrule
\end{tabular}
\end{table}
In this section, we present a detailed view of the device data and distributions for the proposed \ac{IID} and non-\ac{IID} datasets, DESED-FL and URBAN-FL. Table \ref{tab:stats} lists detailed statistics regarding polyphony, i.e. the number of events occurring simultaneously, number and duration of events per sound soundscape. The statistics indicate that the datasets all exhibit similar characteristics, with the only significant different in the max durations between the DESED-FL and URBAN-FL datasets. This is due to the nature of the source datasets. The max duration of the sound events in the URBAN-8K datasets is four seconds, whereas in the DESED-FL source dataset, some events can be longer than four seconds so the max duration is limited to the length of the soundscapes. 

\begin{figure}[H]
    \centering
    \begin{subfigure}[b]{\textwidth}
        \centering
        \includegraphics[width=\textwidth]{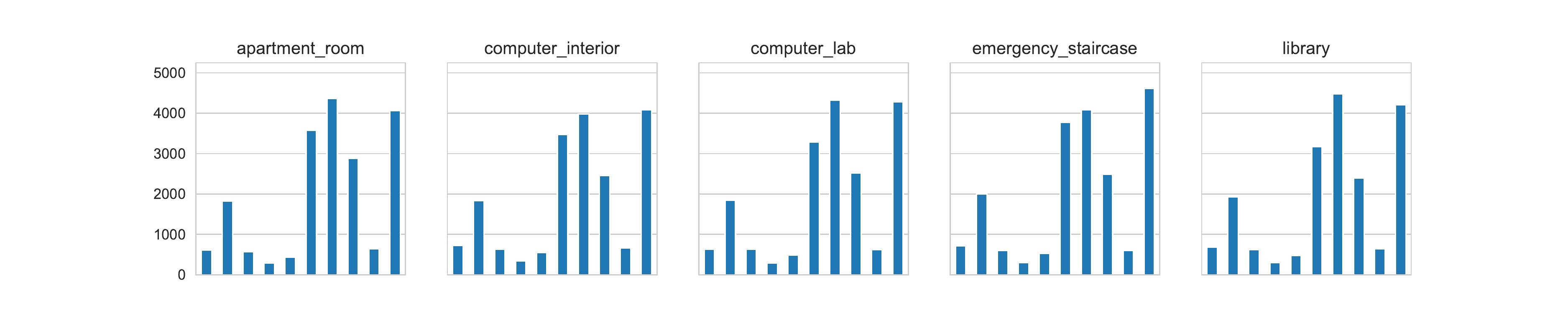}
        \caption{DESED-FL \ac{IID} training data}
        \label{subfig:meandesediid}
    \end{subfigure}%
    \\
    \begin{subfigure}[b]{\textwidth}
        \centering
        \includegraphics[width=\textwidth]{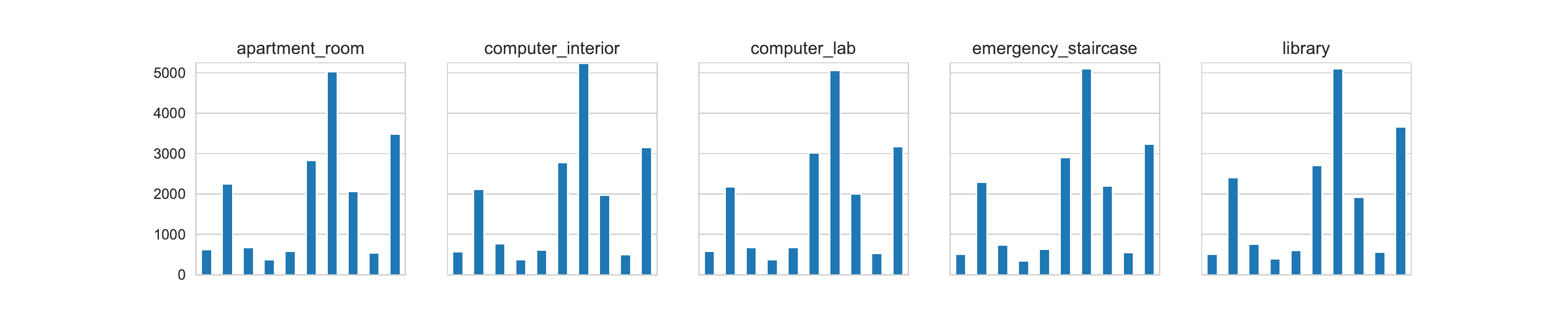}
        \caption{DESED-FL non-\ac{IID} training data}
        \label{subfig:meandesednoniid}
    \end{subfigure}%
    \\
    \begin{subfigure}[b]{\textwidth}
        \centering
        \includegraphics[width=\textwidth]{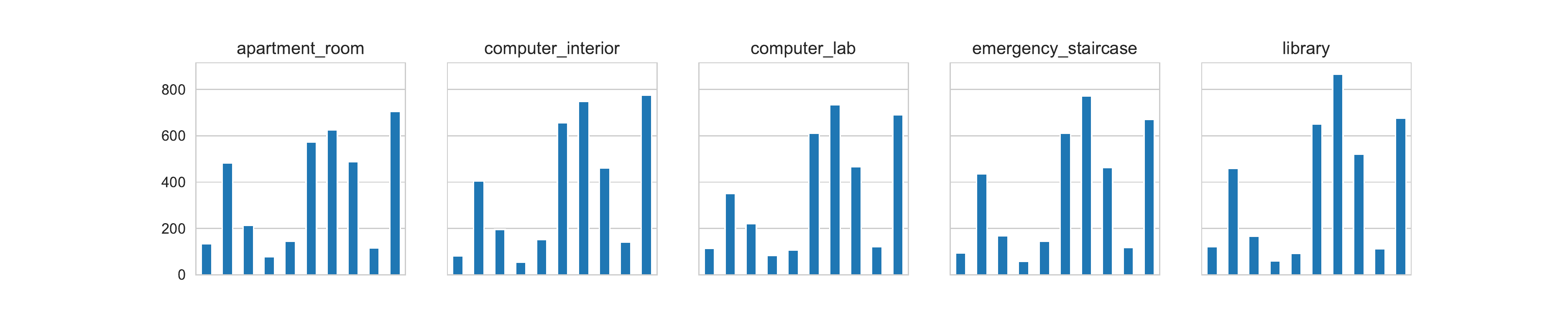}
        \caption{DESED-FL evaluation data}
        \label{subfig:meandesedeval}
    \end{subfigure}%
    \caption{DESED-FL: Total sound event durations for each background for the two versions of training data and the evaluation dataset.  The sound event classes from left to right are: \textit{alarm\_bell\_ringing, blender, cat, dishes, dog, electric\_shaver\_toothbrush, frying, running\_water, speech, and vacuum\_cleaner}.}
    \label{fig:deseddata_means}
\end{figure}

\begin{figure}[H]
    \centering
    \begin{subfigure}[b]{\textwidth}
        \centering
        \includegraphics[width=\textwidth]{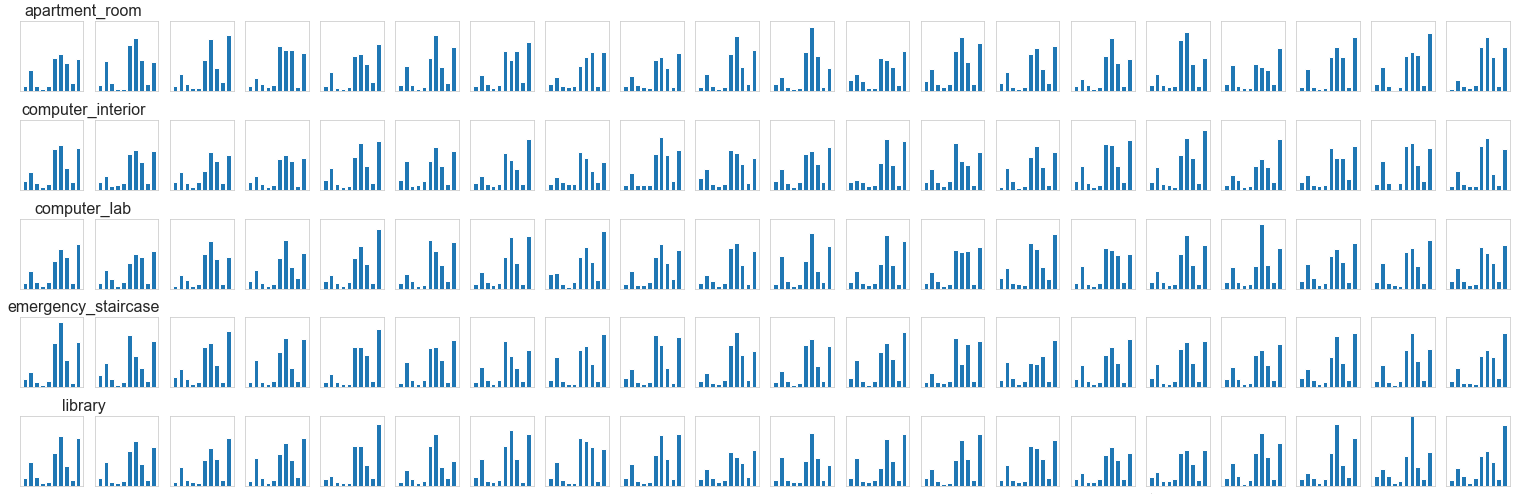}
        \caption{DESED-FL \ac{IID} training data event duration per client}
        \label{subfig:desediid}
    \end{subfigure}%
    \\
    \begin{subfigure}[b]{\textwidth}
        \centering
        \includegraphics[width=\textwidth]{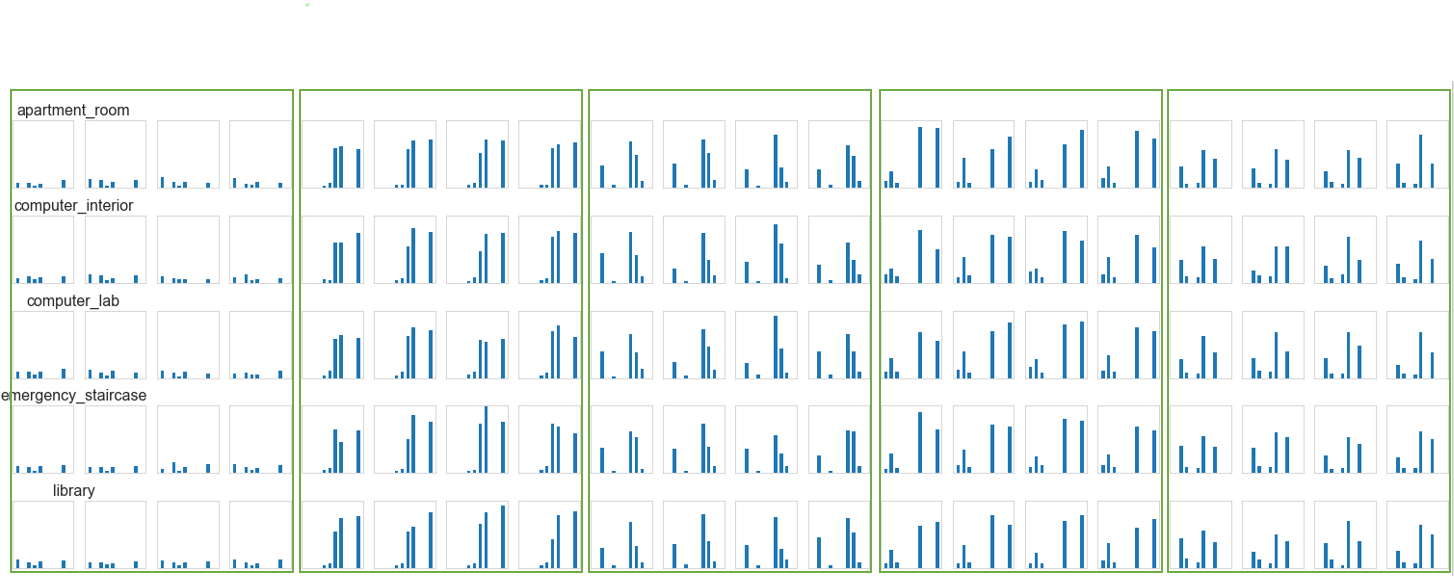}
        \caption{DESED-FL non-\ac{IID} training data event duration per client}
        \label{subfig:desednoniid}
    \end{subfigure}%
    \caption{Distribution of data, by event duration, to client devices for the \ac{DESED}-FL dataset. Each row shows the distribution for a specific background type. The green boxes indicate the five different class distribution groupings for the non-\ac{IID} dataset.}
    \label{fig:data_dists}
\end{figure}

\begin{figure}[H]
    \centering
    \begin{subfigure}[b]{\textwidth}
        \centering
        \includegraphics[width=\textwidth]{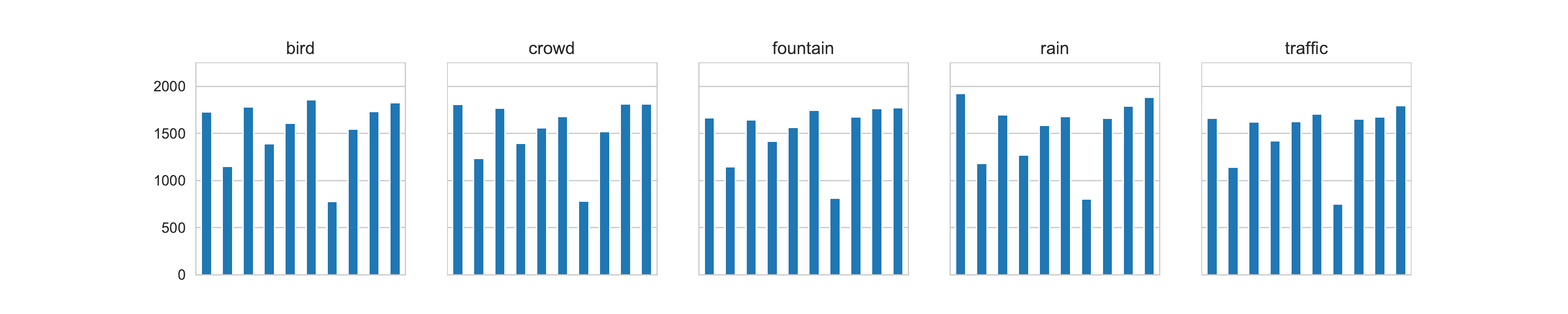}
        \caption{URBAN-FL \ac{IID} training data}
        \label{subfig:meanurbaniid}
    \end{subfigure}%
    \\
    \begin{subfigure}[b]{\textwidth}
        \centering
        \includegraphics[width=\textwidth]{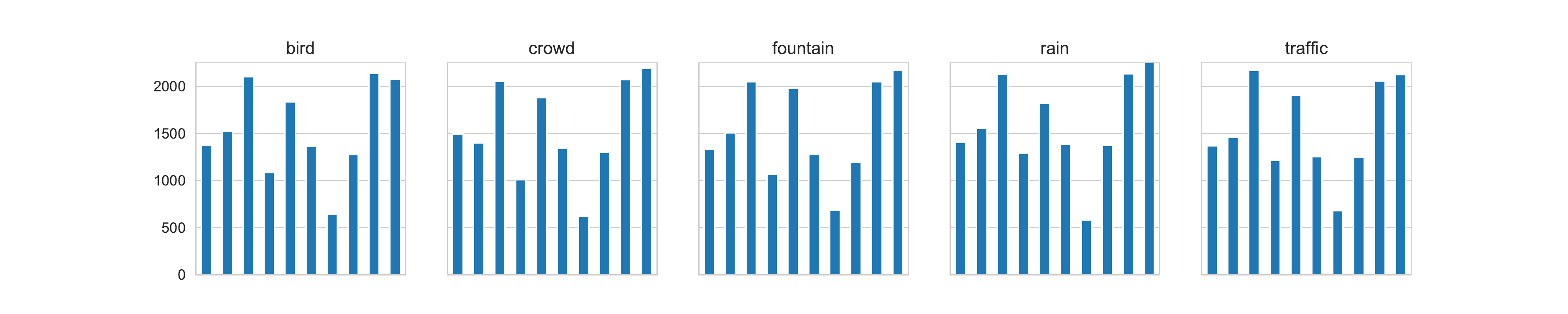}
        \caption{URBAN-FL non-\ac{IID} training data}
        \label{subfig:meanurbannoniid}
    \end{subfigure}%
    \\
    \begin{subfigure}[b]{\textwidth}
        \centering
        \includegraphics[width=\textwidth]{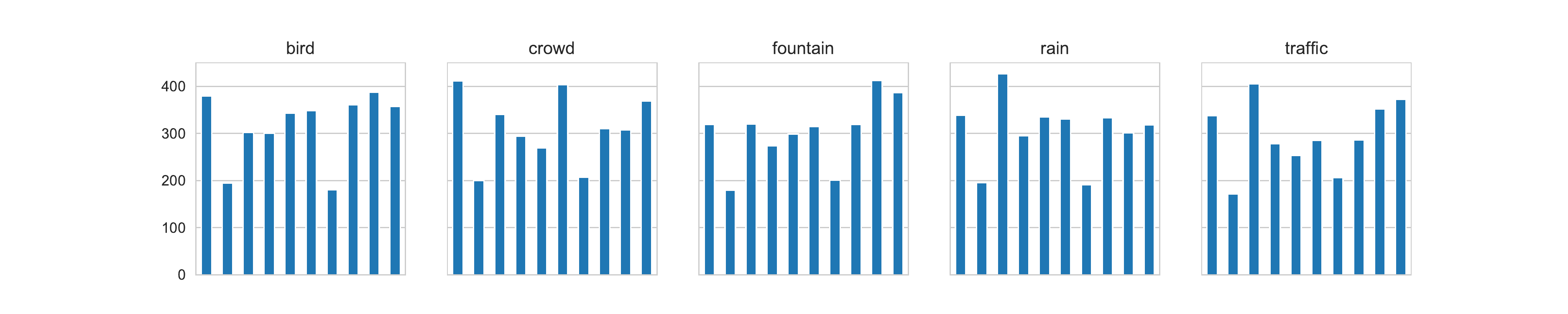}
        \caption{URBAN-FL evaluation data}
        \label{subfig:meanurbaneval}
    \end{subfigure}%
    \caption{URBAN-FL: Total sound event durations for each background for the two versions of training data and the evaluation dataset.  The sound event classes from left to right are: air\_conditioner, car\_horn, children\_playing, dog\_bark, drilling, engine\_idling, gun\_shot, jackhammer, siren, street\_music}
    \label{fig:urbandata_means}
\end{figure}

\begin{figure}[H]
    \centering
    \begin{subfigure}[b]{\textwidth}
        \centering
        \includegraphics[width=\textwidth]{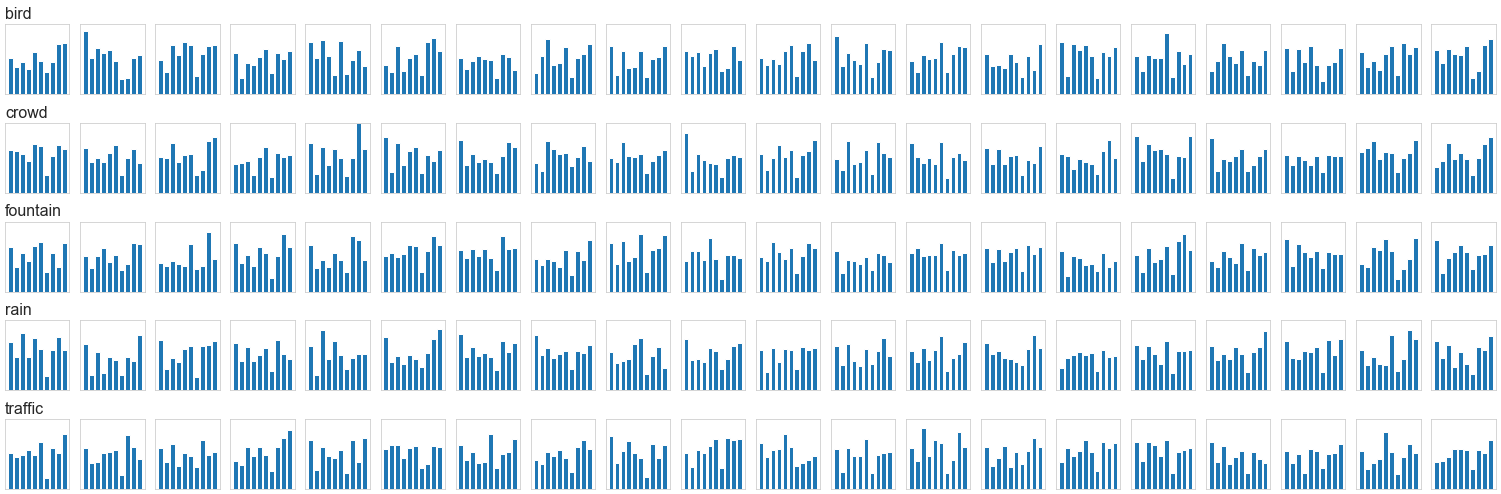}
        \caption{URBAN-FL \ac{IID} training data event duration per client}
        \label{subfig:urbaniid}
    \end{subfigure}%
    \\
    \begin{subfigure}[b]{\textwidth}
        \centering
        \includegraphics[width=\textwidth]{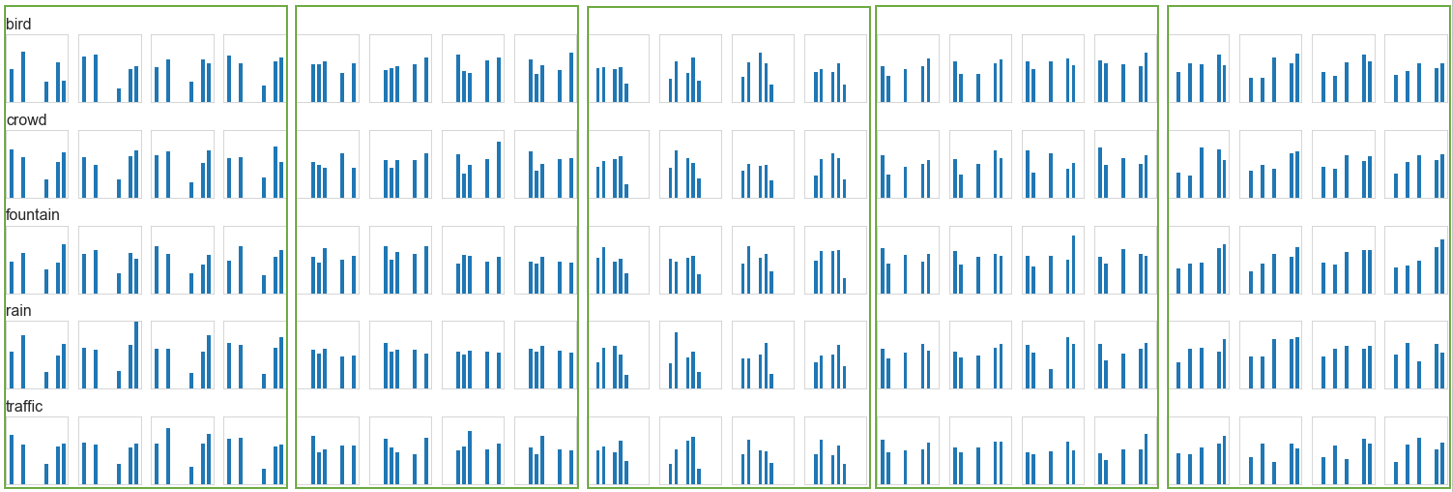}
        \caption{URBAN-FL non-\ac{IID} training data event duration per client}
        \label{subfig:urbannoniid}
    \end{subfigure}%
    \caption{Distribution of data, by event duration, to client devices for the URBAN-FL dataset. Each row shows the distribution for a specific background type. The green boxes indicate the five different class distribution groupings for the non-\ac{IID} dataset.}
    \label{fig:data_dists_urban}
\end{figure}

Figures \ref{fig:deseddata_means} and \ref{fig:urbandata_means} show the total duration of each of the sound events as distributed to the background noise classes for each of the training and evaluation datasets of DESED-FL and URBAN-FL respectively. As expected, the distributions of data between each background. The data was distributed evenly amongst background classes to mitigate any potential background biases in the final trained models. Figures \ref{fig:data_dists} and \ref{fig:data_dists_urban}, provide a detailed look at the individual client distributions. The event durations bar charts for the \ac{IID} data, in Figures \ref{subfig:desediid} and \ref{subfig:urbaniid}, show that the sound events are identically distributed for all devices. Whereas, in the non-\ac{IID} data, shown in Figures \ref{subfig:desednoniid} and \ref{subfig:urbannoniid}, there are five different distributions for each background class, indicated by the surrounding green boxes. This distribution scheme reduces the potential for background bias by ensuring all sound events are distributed among each background type while ensuring non-\ac{IID} client devices. 

\newpage
\section{Detailed Architectures}\label{app:archs}
In this section, we provide details on the three architectures used in our experiments on \ac{AED} with \ac{FL} to foster reproducibility of baseline results on the proposed datasets.  As baseline architectures we propose two standard \ac{CNN} architectures, a small one with close to \num{100000} parameters and a medium-sized one. Additionally, we proprosed a medium-sized \ac{ResNet} based architecture. 

\begin{table}[ht]
\renewcommand{\arraystretch}{1}
\caption{CNN-Base Architecture with \num{542442} parameters}
\label{tab:cnnarch}
\centering
\begin{tabular}{@{} l c c c@{}} \toprule
\textbf{Layer} & \textbf{Output} & \textbf{Kernel Size} & \textbf{Droput}\\\midrule
Conv Block & 16 & (3, 3) & 0.20  \\
Avg Pooling & - & (2, 2) & - \\
Conv Block & 32 & (3, 3) & 0.20  \\
Avg Pooling  & - & (2, 2) & -    \\
Conv Block & 64 & (3, 3) & 0.20  \\
Avg Pooling & - & (1, 2) & -    \\
Conv Block & 128 & (3, 3) & 0.20 \\
Avg Pooling  & - & (1, 2) & -   \\
Conv Block & 128 & (3, 3) & 0.20  \\
Avg Pooling  & - & (1, 2) & -    \\
Conv Block & 128 & (3, 3) & 0.20   \\
Avg Pooling  & - & (1, 2) & -   \\
Conv Block & 128 & (3, 3) & 0.20  \\
Global Avg Pooling & - & - & - \\
Dense & 10 & - & - \\
Sigmoid & - & - & - \\

\end{tabular}
\end{table}

\begin{figure}[htpb]
\centering
    \begin{subfigure}[b]{0.495\textwidth}
    \centering
        \includegraphics[width=\linewidth]{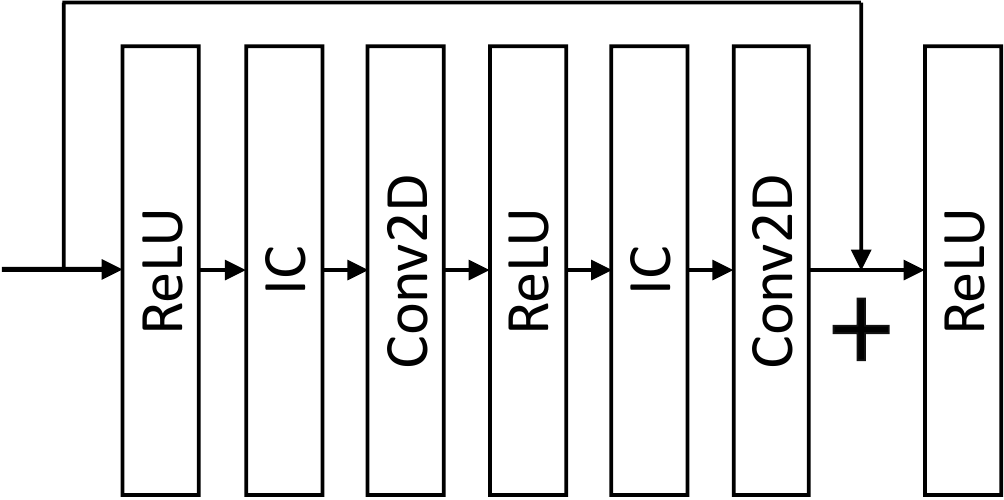}
        \caption{ResNet Block with IC Layers}
        \label{subfig:resnet}
    \end{subfigure} \hfill%
    \begin{subfigure}[b]{0.30\textwidth}
    \centering
        \includegraphics[width=\linewidth]{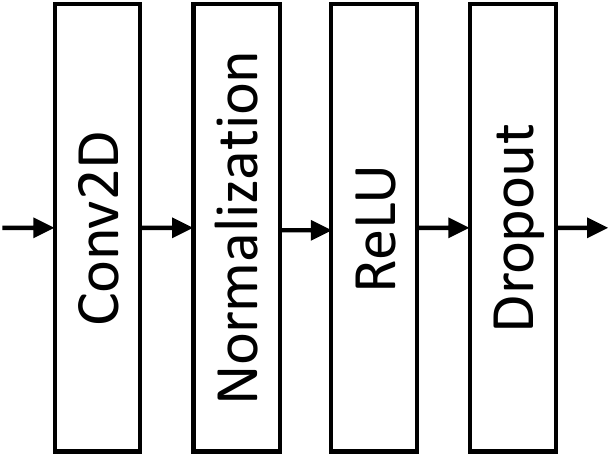}
        \caption{Convolution Block}
        \label{subfig:cnn}
    \end{subfigure}
    \caption{Blocks used for the ResNet and CNN architectures}
    \label{fig:arch_blocks}
\end{figure}

Both CNN-Base and CNN-Small are standard CNN architectures composed of convolution blocks each followed by a pooling layer. Each convolutional block, shown in Figure~\ref{subfig:cnn}, contains a 2D convolution, a normalization layer, an activation layer, and 20\% dropout.  Global pooling is performed before the final output layer.
CNN-Base is based on the baseline architecture from the \ac{DCASE} 2019 challenge for \ac{AED} ~\citeApp{turpault:dcase2019baseline} without the recurrent layers, and is composed of seven convolutional blocks. The recurrent layers were removed in order to maintain compatibility with neuromorphic hardware. The architecture and hyperparameters are listed in Table~\ref{tab:cnnarch}.  The resulting architecture contains \num{542442}~trainable parameters.

\begin{table}[ht]
\renewcommand{\arraystretch}{1}
\caption{\ac{CNN-S} Architecture with \num{115,434} parameters}
\label{tab:cnnsmallarch}
\centering
\begin{tabular}{@{} l c c c@{}} \toprule
\textbf{Layer} & \textbf{Output} & \textbf{Kernel Size} & \textbf{Dropout} \\\midrule
Conv Block & 16 & (3, 3) & 0.2 \\
Pooling 2D & - & (2, 2) & -  \\
Conv Block & 32 & (3, 3) & 0.2 \\
Pooling 2D & - & (2, 2) & -    \\
Conv Block & 64 & (3, 3) & 0.2  \\
Pooling 2D & - & (2, 2) & -    \\
Conv Block & 128 & (3, 3)  & 0.2 \\
Global Pooling & - & - & - \\
Dense & 128 & - & - \\
ReLu & - & - & - \\
Dense & 10 & - & - \\
Sigmoid & - & -  & -\\

\end{tabular}
\end{table}

CNN-Small is a \ac{CNN} that was designed using Bayesian Optimization neural architecture search~\citeApp{Snoek2012:bayes_opt}\footnote{Implemented with \url{http://github.com/fmfn/BayesianOptimization}} to keep the model close to \num{100000}~parameters.  The final architecture is composed of four convolutional blocks, seen in~\ref{subfig:cnn} and a classification network with one hidden layer and one output layer, for a total of \num{115434}~parameters. The architecture and hyperparameters are listed in Table~\ref{tab:cnnarch}.

\begin{table}[H]
\renewcommand{\arraystretch}{1}
\caption{Resnet Architecture with \num{422090} parameters}
\label{tab:resnetarch}
\centering
\begin{tabular}{@{} l c c c c@{}} \toprule
\textbf{Layer} & \textbf{Output} & \textbf{Kernel Size 1} & \textbf{Kernel Size 2}  & \textbf{Dropout} \\\midrule
Conv 2D & 64 & (5, 5) & - & - \\
Relu & - & -  & - & - \\
ResNet Block & 64 & (3, 3) & (1, 1) & 0.10\\
Avg Pooling & - & (2, 2) & - & - \\
ResNet Block & 64 & (3, 3) & (3, 3) & 0.10 \\
Avg Pooling& - & (2, 2) & - & - \\
ResNet Block & 64 & (3, 3) & (3, 3) & 0.10 \\
Avg Pooling & - & (2, 2) & - & - \\
ResNet Block & 128 & (3, 3) & (1, 1) & 0.10 \\
Avg Pooling & - & (2, 2) & - & - \\
ResNet Block & 256 & (1, 1) & (1, 1) & 0.10 \\
Avg Global Pooling & - & - & - & - \\
Dense & 10 & - & - & - \\
Sigmoid & - & - & - & - \\
\end{tabular}
\end{table}

The \ac{ResNet} architecture is composed of \ac{ResNet} blocks with \ac{IC} layers as proposed by Chen et~al.~\citeApp{chen2019.rethinking} shown in Fig.~\ref{subfig:resnet} with hyperparameters inspired by Koutini et~al.~\citeApp{Koutini2019} to minimize the receptive field of the model. The \ac{IC} blocks contain a normalization layer followed by dropout.  For additional regularization, l2 weight regularization is performed as well as disabling of centering and scaling during the normalization process \citeApp{McDonnell2020}. The final architecture contains \num{422090}~trainable parameters.

\newpage
\section{Extended Results}\label{app:res}
\begin{figure}[ht]
\centering
    \begin{subfigure}[b]{\textwidth}
        \centering
        \includegraphics[width=0.25\textwidth]{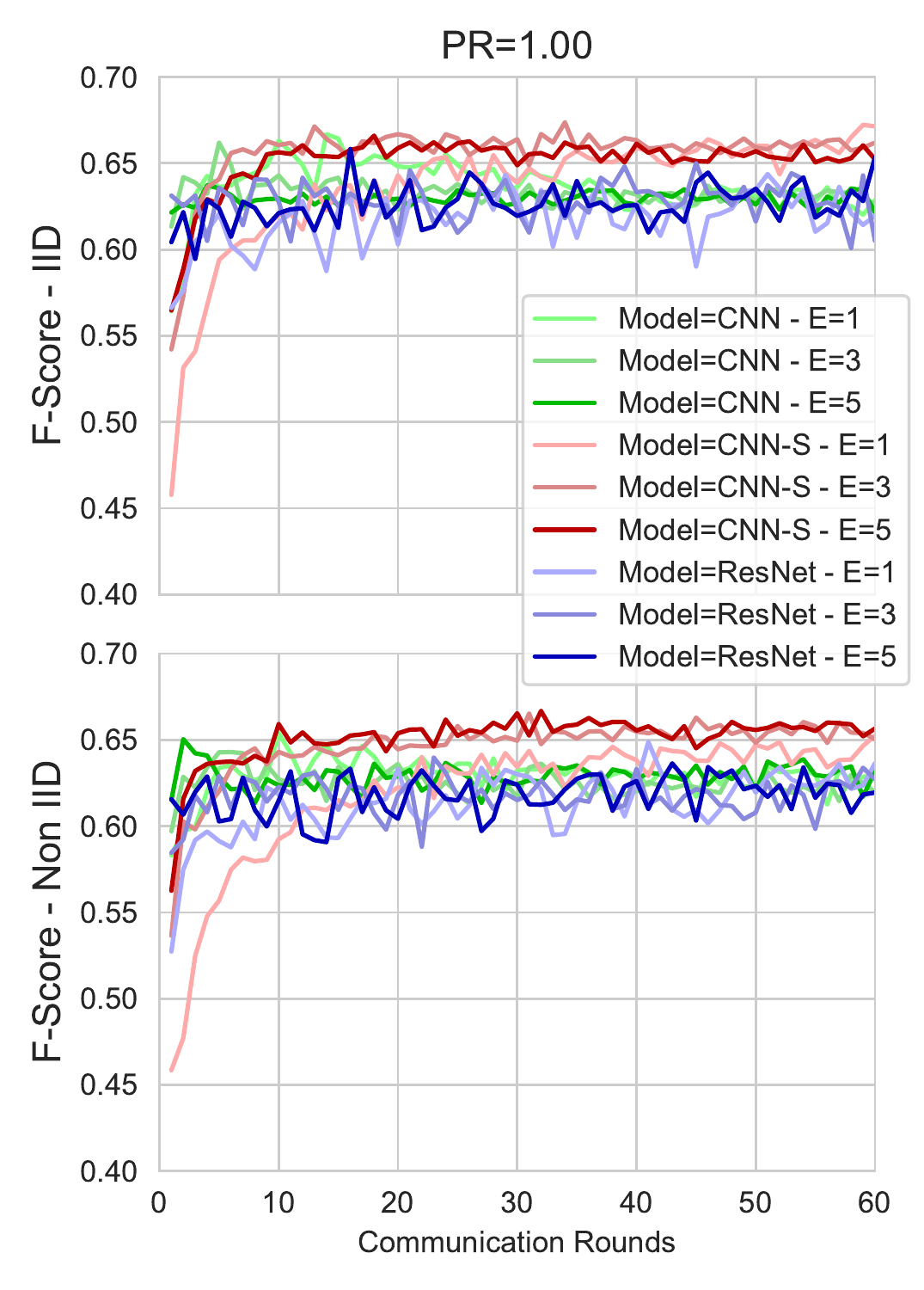}
        \caption{N Clients = 1}
        \label{subfig:results:nc1}
    \end{subfigure} \\

    \begin{subfigure}[b]{\textwidth}
        \centering
        \includegraphics[width=0.75\textwidth]{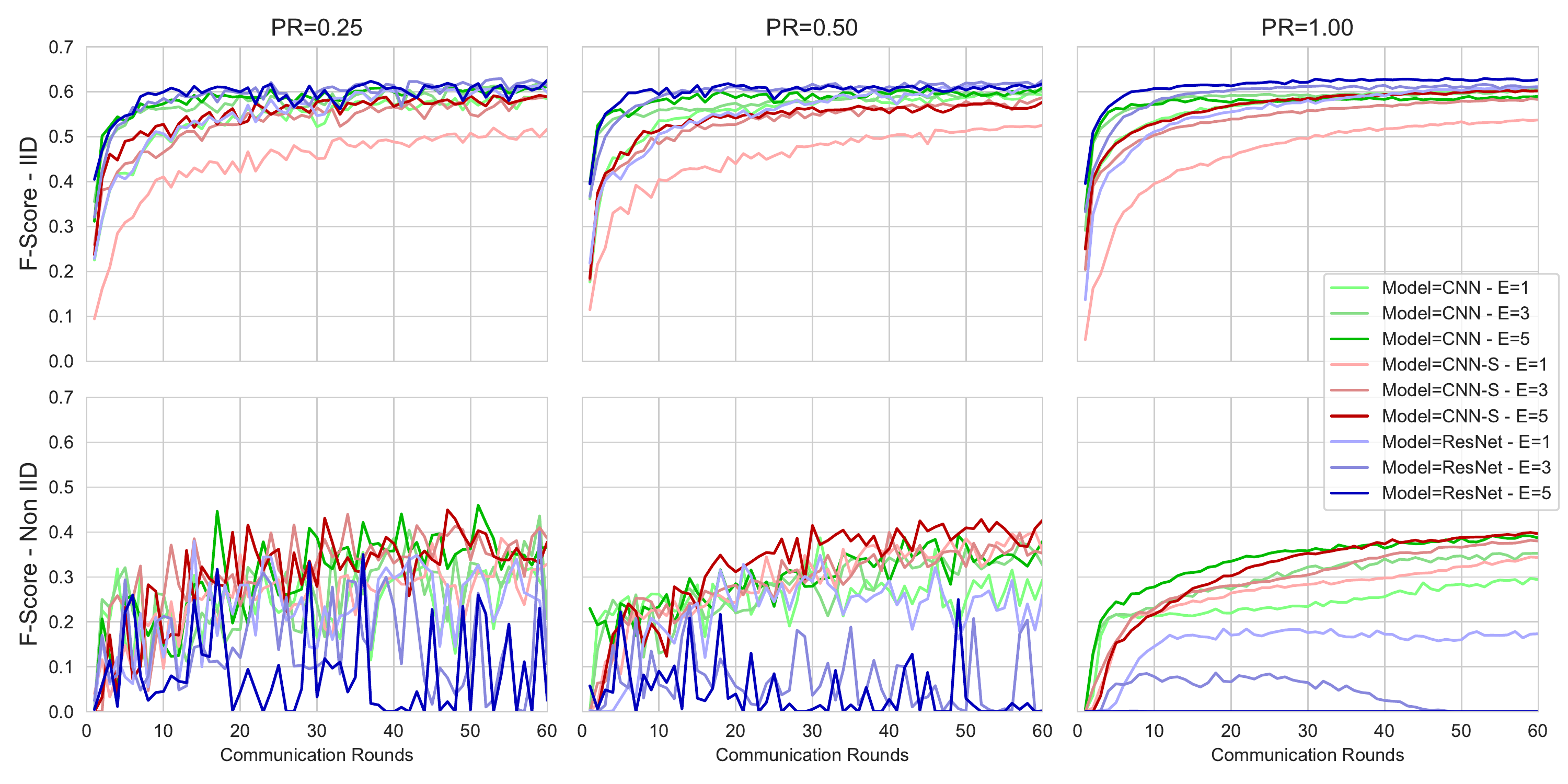}
        \caption{N Clients = 25}
        \label{subfig:results:nc25}
    \end{subfigure} \\
    
        \begin{subfigure}[b]{\textwidth}
        \centering
        \includegraphics[width=0.75\textwidth]{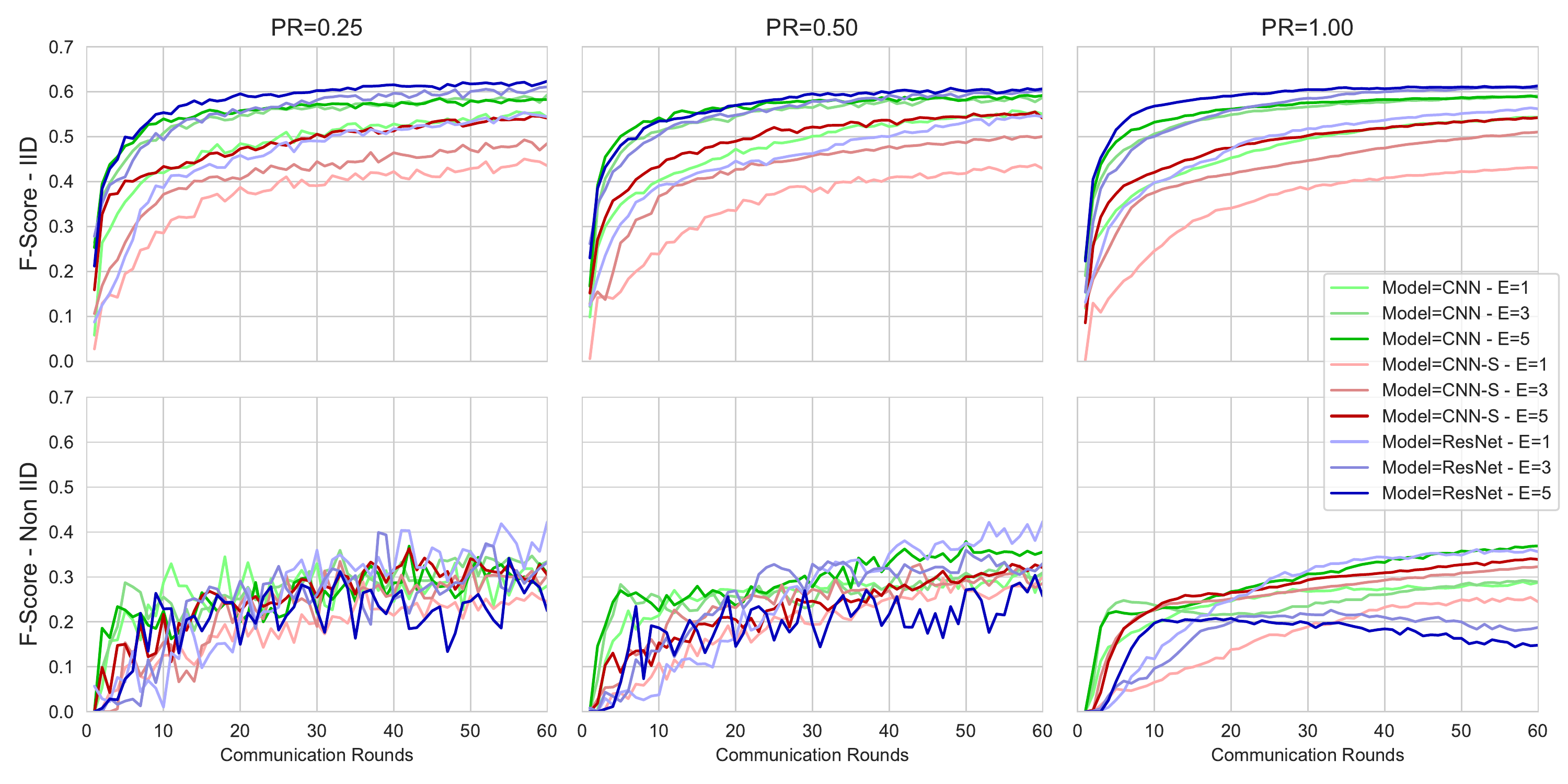}
        \caption{N Client = 100}
        \label{subsubfig:results:nc100}
    \end{subfigure} \\
    
    \caption{DESED-FL: Federated Learning F1 (micro average) Score on Validation Data}
    \label{fig:results}
\end{figure}

\begin{figure}[ht]
\centering
    \begin{subfigure}[b]{\textwidth}
        \centering
        \includegraphics[width=0.25\textwidth]{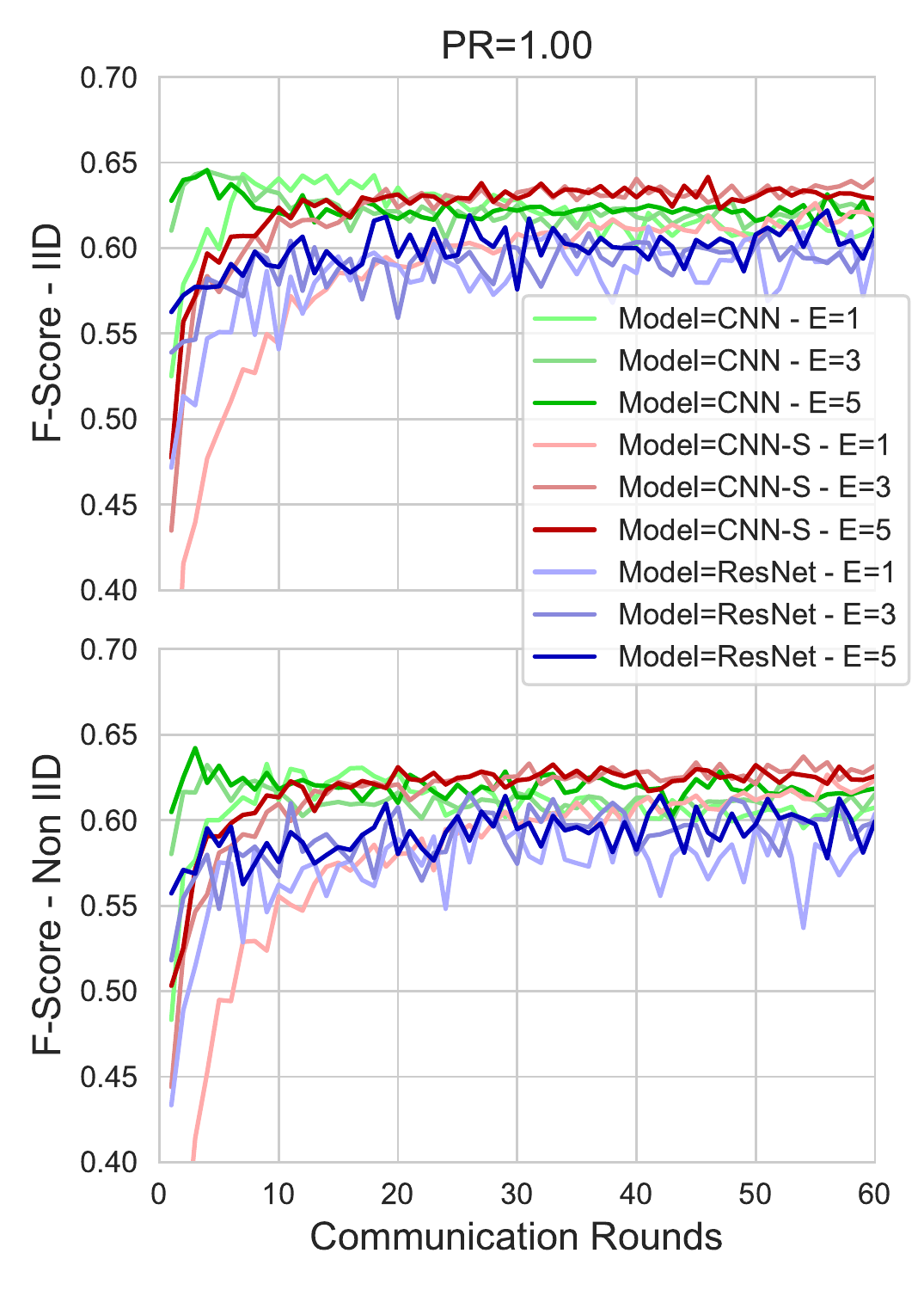}
        \caption{N Clients = 1}
        \label{subfig:urban_results:nc1}
    \end{subfigure} \\

    \begin{subfigure}[b]{\textwidth}
        \centering
        \includegraphics[width=0.75\textwidth]{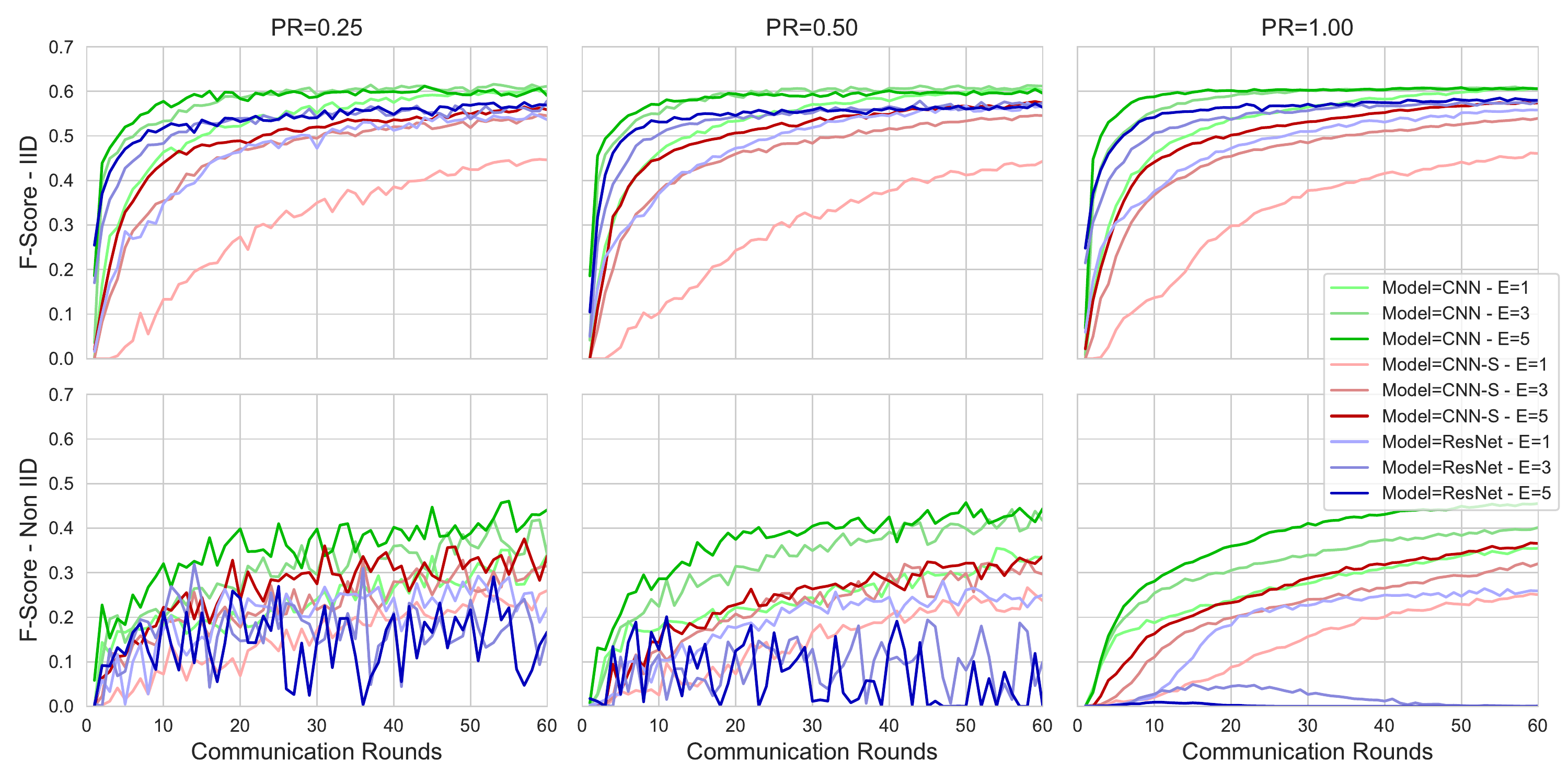}
        \caption{N Clients = 25}
        \label{subfig:urban_results:nc25}
    \end{subfigure} \\
    
        \begin{subfigure}[b]{\textwidth}
        \centering
        \includegraphics[width=0.75\textwidth]{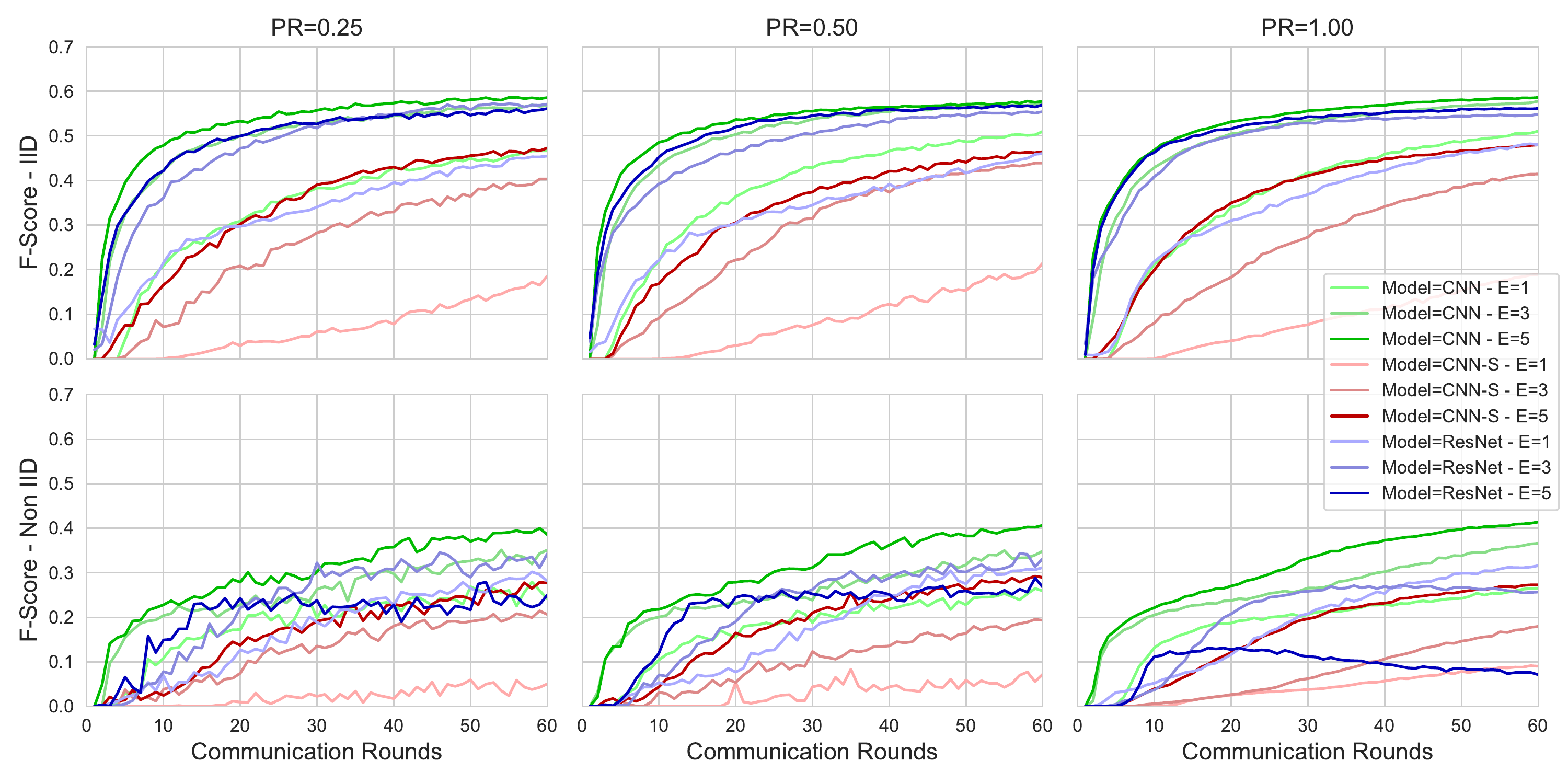}
        \caption{N Client = 100}
        \label{subfig:urban_results:nc100}
    \end{subfigure} \\
    
    \caption{URBAN-FL: Federated Learning F1 (micro average) Score on Validation Data}
    \label{fig:urban_results}
\end{figure}

In this section, we present extended results from our \ac{FL} experiments.  In addition, to $r_p=0.5$, presented in the main article, results for $r_p=\{0.25,\:1.0\}$ are shown in Figures \ref{fig:results} and \ref{fig:urban_results}. The percentage of clients involved in a training round has minimal effects on the training peformance for \ac{IID} data, while in the case of non-\ac{IID} training is more stabilized as the percentage increases. This is most likely a result of the central server receiving weight gradients from all possible distributions at each training round. 

\end{appendices}

\clearpage

% \printbibliography